\documentclass[12pt,preprint]{aastex}
\begin{document}
\setcounter{figure}{0}
\title{Proper Motion of the Draco Dwarf Galaxy Based On
\textit{Hubble Space Telescope} Imaging.\footnote{Based on observations
with the NASA/ESA \textit{Hubble Space Telescope}, obtained at the Space
Telescope Science Institute, which is operated by the Association of
Universities for Research in Astronomy, Inc., under NASA contract NAS
5-26555.  The observations are associated with programs 10229 and
10812.}}

\author{Carlton Pryor}
\affil{Dept. of Physics and Astronomy, Rutgers, the State University
of New Jersey, 136~Frelinghuysen Rd., Piscataway, NJ 08854--8019 \\
E-mail address: pryor@physics.rutgers.edu}

\author{Slawomir Piatek} \affil{Dept. of Physics, New Jersey Institute
of Technology, Newark, NJ 07102 \\
E-mail address: piatek@physics.rutgers.edu}


\author{Edward W.\ Olszewski}
\affil{Steward Observatory, The University of Arizona, Tucson, AZ 85721 \\
E-mail address: eolszewski@as.arizona.edu}

\begin{abstract}

We have measured the proper motion of the Draco dwarf galaxy using
images at two epochs with a time baseline of about two years taken with
the Hubble Space Telescope and the Advanced Camera for Surveys.  Wide
Field Channel 1 and 2 provide two adjacent fields, each containing a
known QSO.  The zero point for the proper motion is determined using
both background galaxies and the QSOs and the two methods produce
consistent measurements within each field.  Averaging the results from
the two fields gives a proper motion in the equatorial coordinate system
of
$(\mu_{\alpha},\mu_{\delta}) = (17.7\pm 6.3, -22.1 \pm 6.3)$~mas~century$^{-1}$
and in the Galactic coordinate system of
$(\mu_{\ell},\mu_{b}) = (-23.1\pm 6.3, -16.3 \pm 6.3)$~mas~century$^{-1}$.
Removing the contributions of the motion of the Sun and of the LSR to
the measured proper motion yields a Galactic rest-frame proper motion of
$(\mu_{\alpha}^{\mbox{\tiny{Grf}}},\mu_{\delta}^{\mbox{\tiny{Grf}}}) =
(51.4\pm 6.3, -18.7 \pm 6.3)$~mas~century$^{-1}$ and
$(\mu_{\ell}^{\mbox{\tiny{Grf}}},\mu_{b}^{\mbox{\tiny{Grf}}}) =
(-21.8\pm 6.3, -50.1 \pm 6.3)$~mas~century$^{-1}$.  The implied space
velocity with respect to the Galactic center is $(\Pi, \Theta, Z) =
(27\pm 14 , 89\pm 25, -212\pm 20)$~km~s$^{-1}$.  This velocity implies
that the orbital inclination is $70^{\circ}$, with a $95\%$ confidence
interval of $(59^{\circ}, 80^{\circ})$, and that the plane of the orbit
is consistent with that of the vast polar structure (VPOS) of Galactic
satellite galaxies.

\end{abstract}

\keywords{galaxies: dwarf spheroidal --- galaxies: individual (Draco) ---
astrometry: proper motion}

\section{Introduction}
\label{intro}

The \textit{Hubble Space Telescope} (HST) has proven to be an excellent
instrument for astrometry and, in particular, for measuring absolute
proper motions of Galactic globular clusters
\citep[\textit{e.g.},][]{mil06}, Galactic satellite galaxies
\citep[\textit{e.g.},][and references therein]{sohn13}, and even the
more distant galaxy M31 \citep{sav12}.  The measurement methods have
been evolving alongside the successive generations of detectors on HST
and the discovery of additional dwarf galaxies in the vicinity of the
Milky Way.  These measurements require the presence of at least one
object in the science field with a known or negligibly small absolute
proper motion that serves as a standard of rest. The published absolute
proper motions of Galactic globular clusters and dwarf galaxies using
HST data use three types of standards of rest: QSOs, resolved compact
background galaxies, and foreground stellar populations.  Each of these
has advantages and disadvantages which are discussed below.

\citet{p02b} reports the first successful measurement of an absolute
proper motion for Fornax using HST data employing
spectroscopically-confirmed QSOs as standards of rest.  The advantage of
using a QSO is that it is typically sufficiently distant that its PSF is
the same as that of a star and yet it is bright enough to be among the
brighter objects in the field. Thus, a single---empirically
derived---effective point spread function \citep[ePSF]{ak00} determines
the locations of stars and the QSO.  However, there are weaknesses to
this method. 1) The scarcity of suitable QSOs, which forces the science
field to be centered on the location of the QSO and not on the highest
stellar surface density of the target galaxy or globular cluster.  2)
The uncertainty in the position of a single object, the QSO, primarily
determines the uncertainty in the measured proper motion, if there are
enough stars of the target (typically more than 20) in the field.  3) If
the QSO is too bright compared to the stars of the target, the exposure
time may be too short to yield a large enough sample of stars to measure
the proper motion.  4) If the QSO is too faint compared to the stars of
the target, the proper motion is poorly determined even if there are
many stars with high S/N in the sample (see item 2 above). 5) Using QSOs
requires considerable preparatory effort to find QSO candidates in a
deep color-magnitude diagram and then spectroscopically confirm them. 6)
If a target has no suitable background QSOs, its absolute proper motion
cannot ever be measured with this method. 7) A typical QSO is bluer than
a typical star, causing their PSFs to potentially differ despite their
point-like images.

As the exposure time increases, an HST image reveals an increasing
number of distant galaxies, some with compact, almost point-like, cores.
Together, these compact galaxies can act as a standard of rest as shown
by \citet{ma08}.  Although the uncertainty in the position of a single
compact-core galaxy is likely to be worse than that of a QSO, an average
position for tens, or even hundreds, of galaxies can yield an
uncertainty that is comparable to or even better than that for a QSO.
\citet{ma08} and \citet{s10} have developed a method of using galaxies
as the standard of rest, which has been applied to derive the proper
motion of, for example, Leo~II \citep{l11}, M31 \citep{sav12}, and Leo~I
\citep{sohn13}. The advantage of this method is a complete freedom in
placing an HST camera field in the target.  This freedom either allows
the second-epoch observations to be paired with already-existing deep
first-epoch observations meant for photometry and stellar population
studies \citep[\textit{e.g.},][]{sohn13}, which is a good use of
existing resources, or allows placing the field at or very close to the
center of the target to maximize the number of member stars. However,
the method has weaknesses. 1) The exposure time per frame must be long,
\citet{sohn13} use about 1,500~s, in order for there to be enough compact-core
galaxies with a high enough S/N to establish an accurate zero point. 2)
Automated identification of compact-core galaxies does not work
efficiently. The process misidentifies as galaxies image artifacts,
close stellar pairs, and superpositions of a star on nebulosity or a
galaxy. Because the method requires a ``clean'' sample, the selection
must be supported by human judgment. 3) The most compact galaxies
(\textit{e.g.}, AGN) may have an ePSF that differs only subtly from that
of a star, so these ideal reference objects may be rejected by both
automated selection criteria and visual inspection. 4) A galaxy is a
resolved object having a unique morphology. Using a stellar ePSF to
determine its ``center of light'' can lead to large systematic and
random errors.  Instead, the position of each galaxy is measured by
constructing and fitting an individual ``template.'' Because each
template is unique, it necessarily has lower S/N than an ePSF determined
from many stars.

Exploiting the Galactic bulge stars along the line of sight to
Sagittarius, \citet{ppo10} derived an absolute proper motion for this
galaxy by measuring its proper motion relative to bulge stars and by
determining the absolute proper motion of the bulge stars using the
absolute proper motion of the Galactic center itself \citep{rb04}.
This method requires a target that is close in projection to the Galactic
center.  Many Galactic globular clusters satisfy this requirement.

The current article derives the absolute proper motion of the Draco
dwarf galaxy from HST data using both QSOs and compact galaxies as
standards of rest and compares the results obtained with the two
methods. Table~1 lists those properties of Draco that were used in
deriving its space motion.

The rest of this article is organized as follows.
Section~\ref{sec:data} describes the observations and data and
Section~\ref{sec:reduct} explains the process of deriving the proper
motion.  Section~\ref{sec:res} presents results and
Section~\ref{sec:summary} contains a summary and discussion.

\section{Observations and Data}
\label{sec:data}

We obtained first-epoch images for three distinct pointings in the
direction of Draco using the Advanced Camera for Surveys (ACS) and
Wide Field Channel (WFC) combination on HST in cycle 13. For each
pointing, WFC1 is centered on a spectroscopically-confirmed QSO. Ideally,
second-epoch imaging would be obtained with the same detector.
However, only one pointing was imaged in cycle 15 before the failure of
ACS; the failure forced the other two to be imaged with the Wide Field
and Planetary Camera 2 (WFPC2).  Table 2 summarizes the observations,
with column 1 giving the name of the pointing and columns 2 and 3 the
celestial coordinates of the QSO.  The dates of the observations are
in column 3: the upper one is for the first epoch and the lower for
the second; their difference is the time baseline.  It is about
3~years for Dra~1 and Dra~3 and about 2~years for Dra~2.  The final
three columns list the detector, filter, and exposure time.

The second-epoch data taken with PC2 have a smaller field of view and
worse charge-transfer efficiency (CTE) than the first-epoch data taken
with ACS.  The Dra~1 and Dra~3 fields have 16 and 9 stars with $S/N >
20$, respectively, on the PC chip with the QSO. These numbers of stars
are, at best, only marginally sufficient to derive a proper motion. 
Another problem is that the correction for the effects of the degrading
CTE of WFPC2, which produces shifts in the positions of stars comparable
to those from the proper motions \citep{bpp05}, must also be derived
from these few stars.  Thus, obtaining reliable proper motions for these
fields proved impossible. \citet{kmb13} similarly decided that their
WFPC2 data were not useful for measuring the proper motion of the
Magellanic Clouds.  The data from Dra~1 and Dra~3 are not considered
further in this article, though we note that obtaining a second epoch
for these pointings with the repaired ACS would likely reduce the
uncertainty in the proper motion of Draco by more than a factor of five.

\begin{figure}[t!]
\centering
\includegraphics[angle=0,scale=0.47]{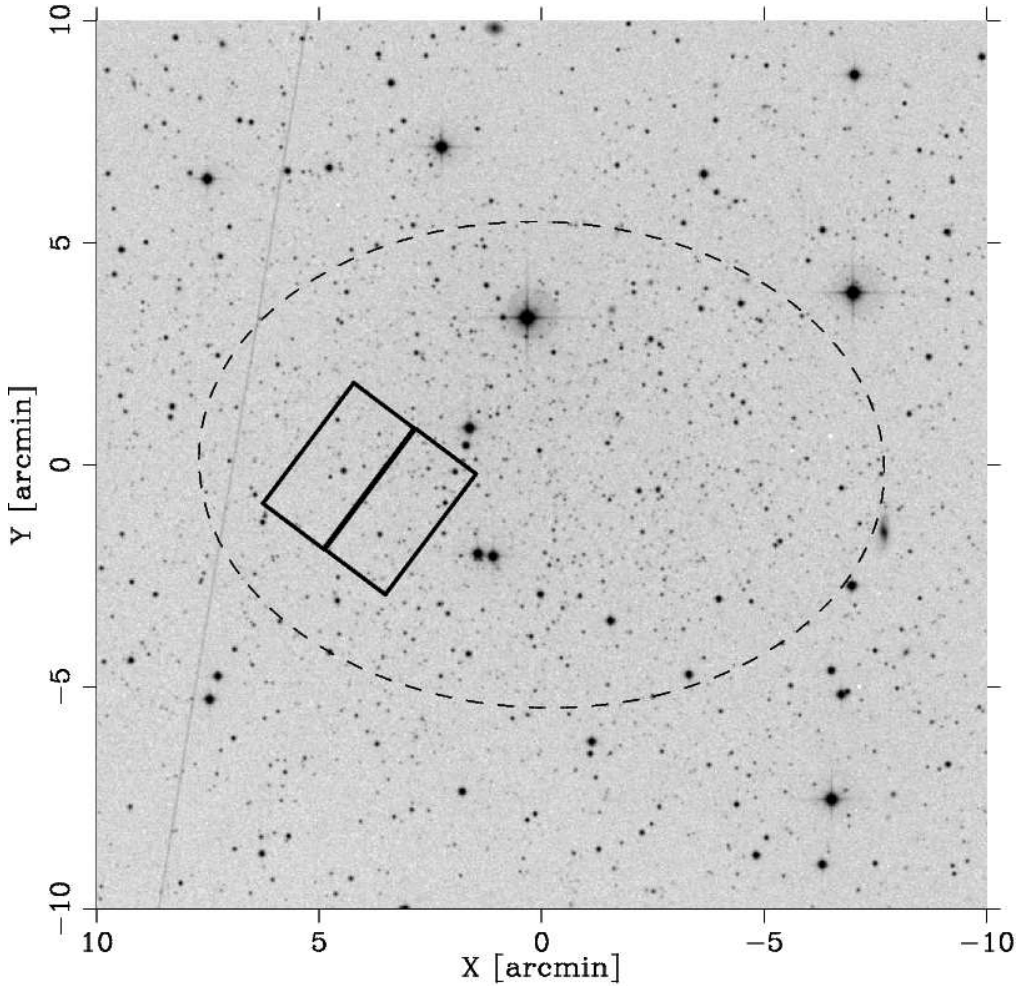}
\caption{Locations of the WFC1 and WFC2 fields
of the Dra~2 pointing on a 20$\times$20~arcmin section of
an image of the sky centered on Draco.  North is up and
East is to the left. The most eastward field is WFC1. The dashed ellipse
delineates the core of Draco. The values for Draco's center, position
angle, ellipticity, and core radius are from Table~1.}
\label{fig:dra2}
\end{figure}

\begin{figure}[t!]
\centering
\includegraphics[angle=0,scale=0.39]{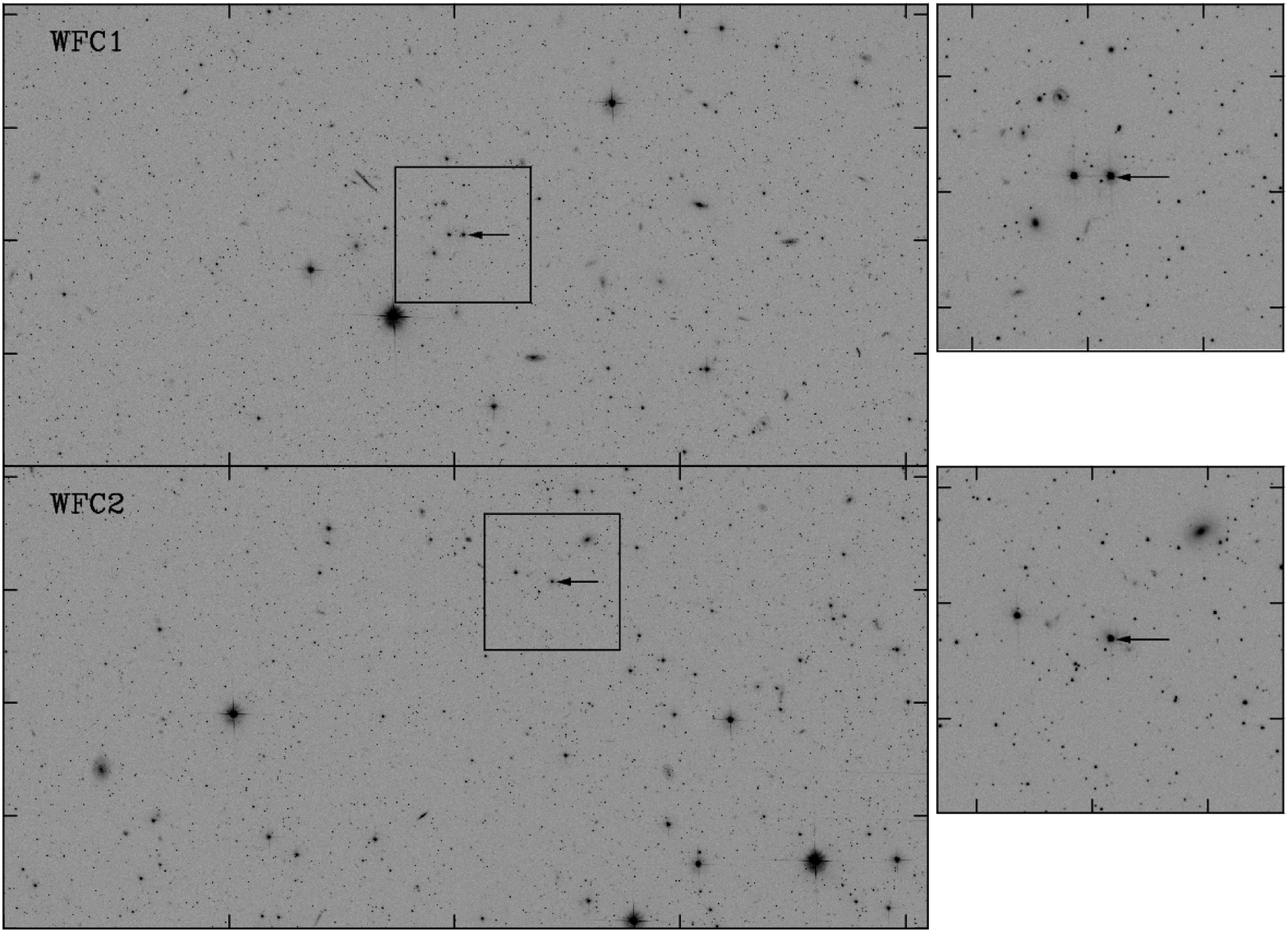}
\caption{Top-left panel: the average of the 19 images taken at the first
epoch, with cosmic rays rejected, for the WFC1 field. The arrow points
to the QSO, which is at the center of the $600\times600$~pixel$^{2}$
box. In the chronologically first exposure, the QSO is at
$(x,y)=(2039.60,1025.04)$~pixel. The smaller top-right panel depicts the
region of the image within the box; again, the arrow points to the QSO.
The bottom two panels are analogous images for the WFC2 field. 
Examination of candidate compact galaxies revealed the presence of a
likely QSO at $(x,y)=(2432.53,1536.54)$~pixel.}
\label{fig:wfc}
\end{figure}

Figure~\ref{fig:dra2} shows the locations of the WFC1 and WFC2 fields of
the Dra~2 pointing on a 20$\times$20~arcmin section of the sky from the
STScI Digitized Sky Survey\footnote{The Digitized Sky Surveys were
produced at the Space Telescope Science Institute under U.S. Government
grant NAG W-2166. The images of these surveys are based on photographic
data obtained using the Oschin Schmidt Telescope on Palomar Mountain and
the UK Schmidt Telescope. The plates were processed into the present
compressed digital form with the permission of these institutions.  The
Second Palomar Observatory Sky Survey (POSS-II) was made by the
California Institute of Technology with funds from the National Science
Foundation, the National Geographic Society, the Sloan Foundation, the
Samuel Oschin Foundation, and the Eastman Kodak Corporation.}. The
dashed ellipse delineates the core of Draco.
The top-left panel of Figure~\ref{fig:wfc}\ is the average, with cosmic
rays rejection, of the 19 images of the WFC1 field taken at the first
epoch.  The QSO is at the center of a $600\times600$~pixel$^{2}$ box and
an arrow points at it. The smaller top-right panel depicts the region of
the image within the box, with the arrow again pointing at the QSO.
Table~2 lists the coordinates of the QSO, which was discovered by
\citet[][their object 264]{k08}.  It has $V = 19.86$ and $V-I = 0.40$
\citep{k08}, $g = 20.29$ and $g-r=0.29$ \citep[][SDSS DR9]{dr9}, and a
redshift of 0.9465 (H.\ C.\ Harris, private communication). Similarly,
the bottom two panels are for the WFC2 field.  The arrow points to an
object which we identified as a likely QSO on the basis of visual
inspection, photometry \citep{r09}, and association with an X-ray source
\citep{fl10}. It is at $(\alpha,\delta)=(17:20:43.10, +57:54:43.0)$
(J2000.0) and has $g = 20.93$ and $g-r=0.49$ \citep[][SDSS DR9]{dr9}.

\section{Data Reduction and Analysis}
\label{sec:reduct}

\subsection{Measuring Coordinates}
\label{sec:mco}

Deriving a proper motion of a resolved stellar system such as Draco
requires accurate measurements of the coordinates of both the member
stars and the zero-point objects --- here QSOs and compact galaxies. 
The following are the steps in measuring these coordinates.

\begin{enumerate}
\item Correct all of the exposures for the degrading charge transfer
efficiency using the standard pipeline processing of ACS images, which
eliminates the need for any further corrections for this effect.

\item Determine the first estimate of an object's coordinates and
magnitude using the stand-alone software package DOLPHOT \citep{dol}. 
There are 19 dithered exposures at each epoch and, thus, this step gives
19 locations for an object --- one per exposure --- in the coordinate
system of the CCD.  Because the exposures are dithered, the locations
can differ by several to a few tens of pixels.

\item Match the same objects across the exposures, determine the
coefficients of the most general quadratic transformation between the
first exposure and each of the subsequent exposures, and transform the
coordinates from those subsequent exposures to the coordinate system
of the first exposure.  These tasks are all done using the DAOPHOT
software package \citep{st87,st92,st94}. From now on in this
article, the ``fiducial coordinate system'' refers to that of the
chronologically first exposure in an epoch.

\item Construct a combined image with higher S/N and no cosmic rays or
hot pixels from the 19 exposures and the coordinate transformations
using only integer-pixel shifts.  The combined image has the same
location and number of pixels as the fiducial image.

\item Discriminate between stars and compact galaxies in both the WFC1
and WFC2 fields using the FWHM of objects found by SExtractor
\citep{ba96} in the combined image.  For every object with a FWHM
greater than 2.5 pixel --- a likely galaxy --- and a S/N greater than
15, construct a sub-sampled image on a 65$\times$65 array of a
13$\times$13~pixel$^2$ cutout using the same methods described by
\citet{p02b}.  Visually inspect each sub-sampled image to reject objects
that are close pairs of stars, stars superimposed on nebulosity or a
galaxy, and galaxies that are too extended to
yield accurate locations.

\item Construct a stellar ePSF that is independent of location in the
image and averaged over all exposures in both epochs using stars with a
S/N greater than 40.  \citet{ak00} and in \citet{p02b} give the
specifics of the construction.

\item Construct an individual template for each identified galaxy and
QSO that is averaged over all exposures in both epochs using a method
similar to that described in \citet{ma08} and \citet{sav12}.  The
template is a two-dimensional function that represents the distribution
of light in the galaxy convolved with the ePSF.  The initial center of
the template is the brightest pixel of the galaxy, which is not
necessarily the geometric center of its image.  The ePSFs for stars and
templates for galaxies are determined by interpolation in a $25 \times
25$ grid of values for a $5 \times 5$-pixel$^2$ region.

\item Measure the location in every exposure for every star and galaxy
using the generated ePSFs and templates.  Correct the coordinates for
the known geometric distortion \citep{a06}, derive the most general
linear transformation between the fiducial coordinate system and those
of the remaining exposures of an epoch, and average the transformed
coordinates at each epoch to produce a final location and uncertainty. 
At the end of this step an object has two sets of coordinates, one in
the fiducial coordinate system of each epoch.

\end{enumerate}

\begin{figure}[b!]
\centering
\includegraphics[angle=-90,scale=0.7]{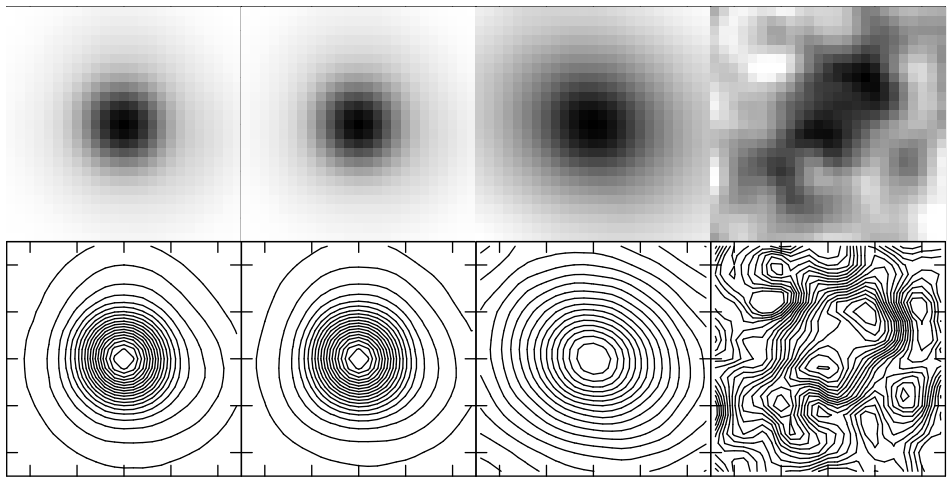}
\caption{Gray-scale maps (top row of panels) and the corresponding
constant-brightness contours (bottom row of panels) of, from left to
right, the average stellar ePSF and the templates of the QSO, the
brightest galaxy, and the faintest galaxy in the WFC1 field.}
\label{fig:WFC1_4psf}
\end{figure}

\begin{figure}[t!]
\centering
\includegraphics[angle=-90,scale=0.7]{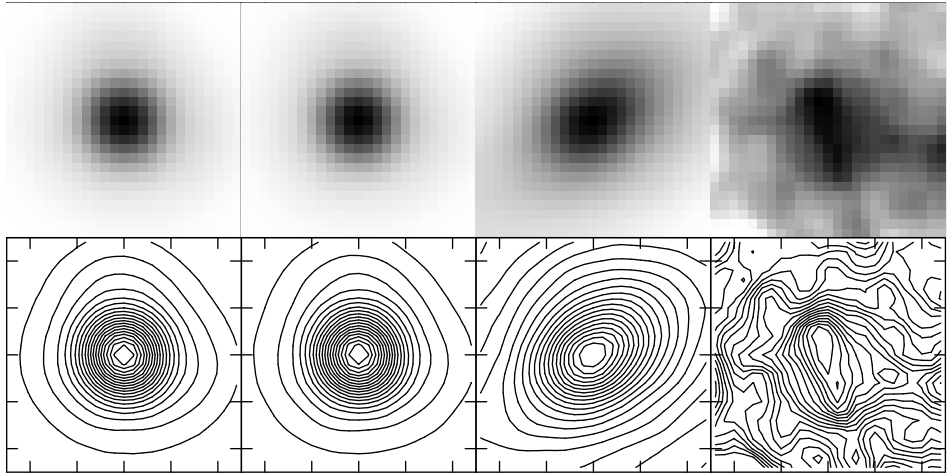}
\caption{The same as Figure~\ref{fig:WFC1_4psf}\ for the WFC2 field.}
\label{fig:WFC2_4psf}
\end{figure}

The above procedure produces coordinates for the set of objects common
to the two epochs: the QSO, 97 selected galaxies, and 1840 stars with
S/N greater than 10 in the WFC1 field and the QSO, 82 selected galaxies,
and 2209 stars with S/N greater than 10 in the WFC2 field.
Figure~\ref{fig:WFC1_4psf}\ depicts for the WFC1 field the gray-scale
maps (top panels) and their corresponding constant-brightness contours
(bottom panels) of, from left to right, the average stellar ePSF and the
templates of the QSO, the brightest galaxy ($\textrm{S/N} = 196$,
$\textrm{FWHM} = 5.5$~pixel), and the faintest galaxy
($\textrm{S/N}=9.4$, $\textrm{FWHM}=6.3$~pixel).  Similarly,
Figure~\ref{fig:WFC2_4psf}\ depicts for the WFC2 field, from left to
right, the average stellar ePSF and the templates of the QSO, the
brightest galaxy ($\textrm{S/N}=138$, $\textrm{FWHM}=4.0$~pixel), and
the faintest galaxy ($\textrm{S/N}=9.9$, $\textrm{FWHM}=4.9$~pixel).
Note that the above values for the S/N are calculated only using pixels
in the $5\times 5$~pixels$^2$ array, while the values for the FWHM are
calculated by Sextractor for the entire galaxy and may not measure the
width of compact core.

Figures~\ref{fig:WFC1_4psf} and \ref{fig:WFC2_4psf} show that, for both
fields, the template of the QSO is nearly identical to the stellar ePSF.
Thus, the results reported in the rest of this article are derived by
fitting the stellar ePSF to the images of the QSOs.  This approach
allows a direct comparison between the technique that measures a proper
motion using a QSO as the astrometric zero-point, such as \citet{p02b},
and that which uses galaxies.  The proper motions derived from the QSOs
by fitting their templates do not differ significantly from those
derived by fitting the ePSf in either value or the size of the
uncertainty.

\begin{figure}[t!]
\centering
\includegraphics[angle=-90,scale=0.72]{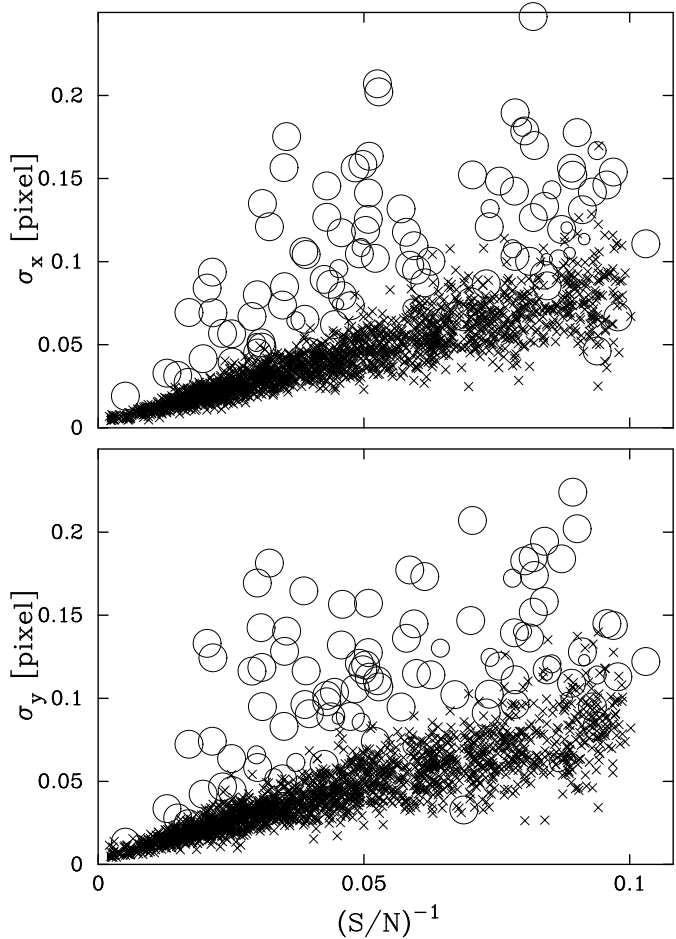}
\caption{The rms scatter around the mean of the X-component (top panel)
and the Y-component (bottom panel) of the centroid as a function of
$(\textrm{S/N})^{-1}$ for the first epoch exposures and WFC1 field. The
slanted crosses are for stars and open circles for galaxies, with the
size of the circle proportional to the FWHM.}
\label{fig:WFC1_uncs}
\end{figure}

\begin{figure}[t!]
\centering
\includegraphics[angle=-90,scale=0.72]{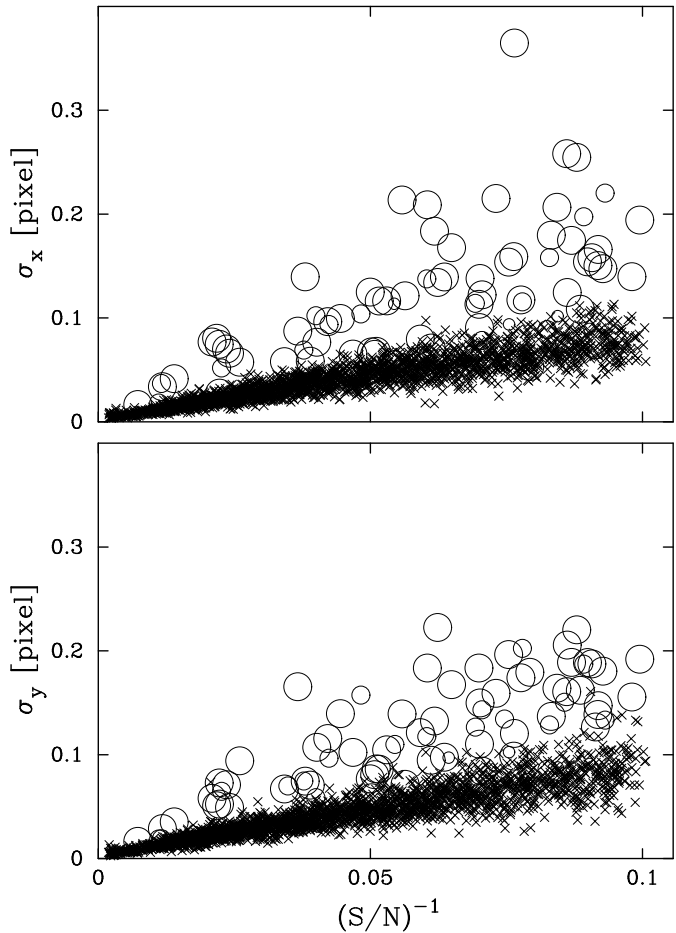}
\caption{The same as Figure~\ref{fig:WFC1_uncs}, except with a
different vertical scale, for the WFC2 field.}
\label{fig:WFC2_uncs}
\end{figure}

        The uncertainty in the location of an object, as estimated
from the scatter of the transformed coordinates around the mean,
depends approximately linearly on $(\textrm{S/N})^{-1}$, as shown in
Figure~\ref{fig:WFC1_uncs}\ and ~\ref{fig:WFC2_uncs} for the WFC1 and
WFC2 fields, respectively.  The slanted crosses in the figures are
for stars and the open circles are for galaxies, with the size of the
circle proportional to the FWHM.  The plots show that the positional
uncertainty for the stars with the highest S/N is about 0.01~pixel and this
increases to 0.05~pixel at a S/N of 20.  Galaxies have larger uncertainties
than stars at a given S/N, as expected from their larger FWHMs.

\subsection{Measuring the Proper Motion}
\label{sec:mpm}

        The procedure for measuring the proper motion of Draco used in
this paper is similar to that in our previous work
\citep[e.g.,][]{p02b}.  In that work, the stars of the galaxy determine
a comoving coordinate system in the reference frame of the first epoch
that we call the standard coordinate system.  Average coordinates in the
fiducial system of the second epoch are transformed into the standard
system using the most general quadratic transformation, which in the
current work needed to be supplemented by adding the term $x^3$ to
eliminate an obvious trend in the residuals.  In the standard coordinate
system, the stars of the galaxy are at rest, whereas the QSO (and field
stars) are moving.  The reverse of the motion of the QSO in the standard
coordinate system is the proper motion of the galaxy.

	In the current work, background galaxies are treated like the QSO
was previously and, thus, the proper motion of Draco is the reverse of
the average motion of the galaxies in the standard coordinate system.
Although the background galaxies are numerous enough to determine a
fixed coordinate system at both epochs, the more numerous and
better-measured stars of Draco determine a more accurate standard
coordinate system.  In both the previous and current work it is
necessary to increase the uncertainties of the average coordinates at
the two epochs to produce a chi-square ($\chi^2$) of one for the
transformation of the coordinates between the two epochs. The current
data favor making this increase by adding a constant in quadrature to
the uncertainties over multiplying them by a constant, as was done in
our previous work.  The additive constant is 0.0048~pixel for the WFC1
field and 0.0046~pixel for the WFC2 field.

	The standard coordinate system is co-moving with Draco. Let $\mu_x$
and $\mu_y$ be the $x$ and $y$ components, respectively, of the motion
(in pixel~yr$^{-1}$) of an object in this coordinate system. Thus, a
star of Draco should have $(\mu_x, \mu_y)$ consistent with zero, whereas
the field stars, background galaxies, and the QSOs have a non-zero
motion.  It is possible for a field star to have $(\mu_x,
\mu_y)=(0,0)$~pixel~yr$^{-1}$ if its proper motion is the same as that
of Draco, though this is an unlikely occurrence.  The set of stars that
determine the standard coordinate system results from iteratively
rejecting stars whose $\mu_x$ or $\mu_y$ differ from zero by a
statistically significant amount. 

\begin{figure}[p]
\centering
\includegraphics[angle=-90,scale=0.68]{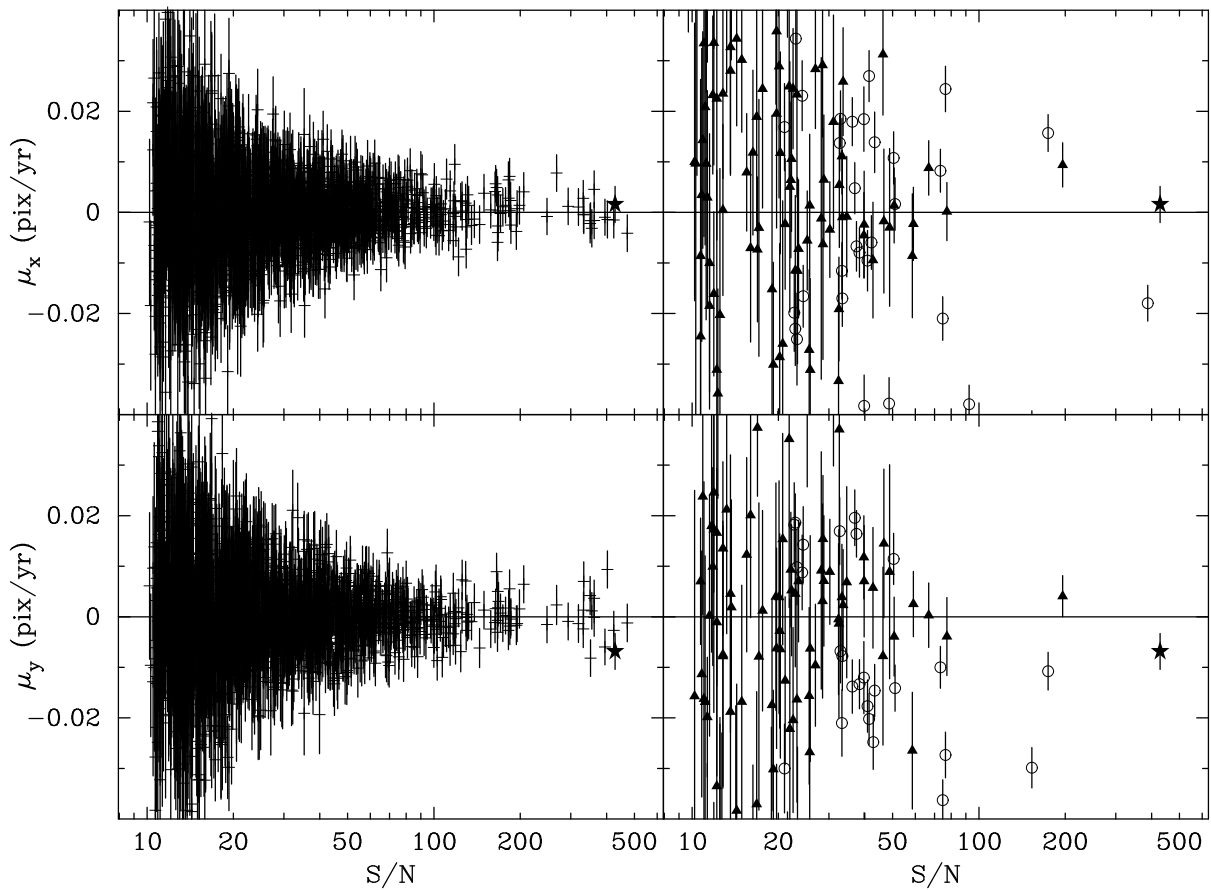}
\caption{Motion in the standard coordinate system, $\mu_{x}$ and $\mu_{y}$, 
in pixel~yr$^{-1}$ as a function of S/N for the WFC1 field. The panels on the
left are for the likely stars of Draco (plus symbols) and the QSO
(filled star), while those on the right are for the galaxies (filled
triangles), field stars (open circles), and - again - the QSO.  Only
field stars with S/N greater than 20 are shown.
}
\label{fig:mu_xy_SN_WFC1}
\end{figure}

\begin{figure}[p]
\centering
\includegraphics[angle=-90,scale=0.68]{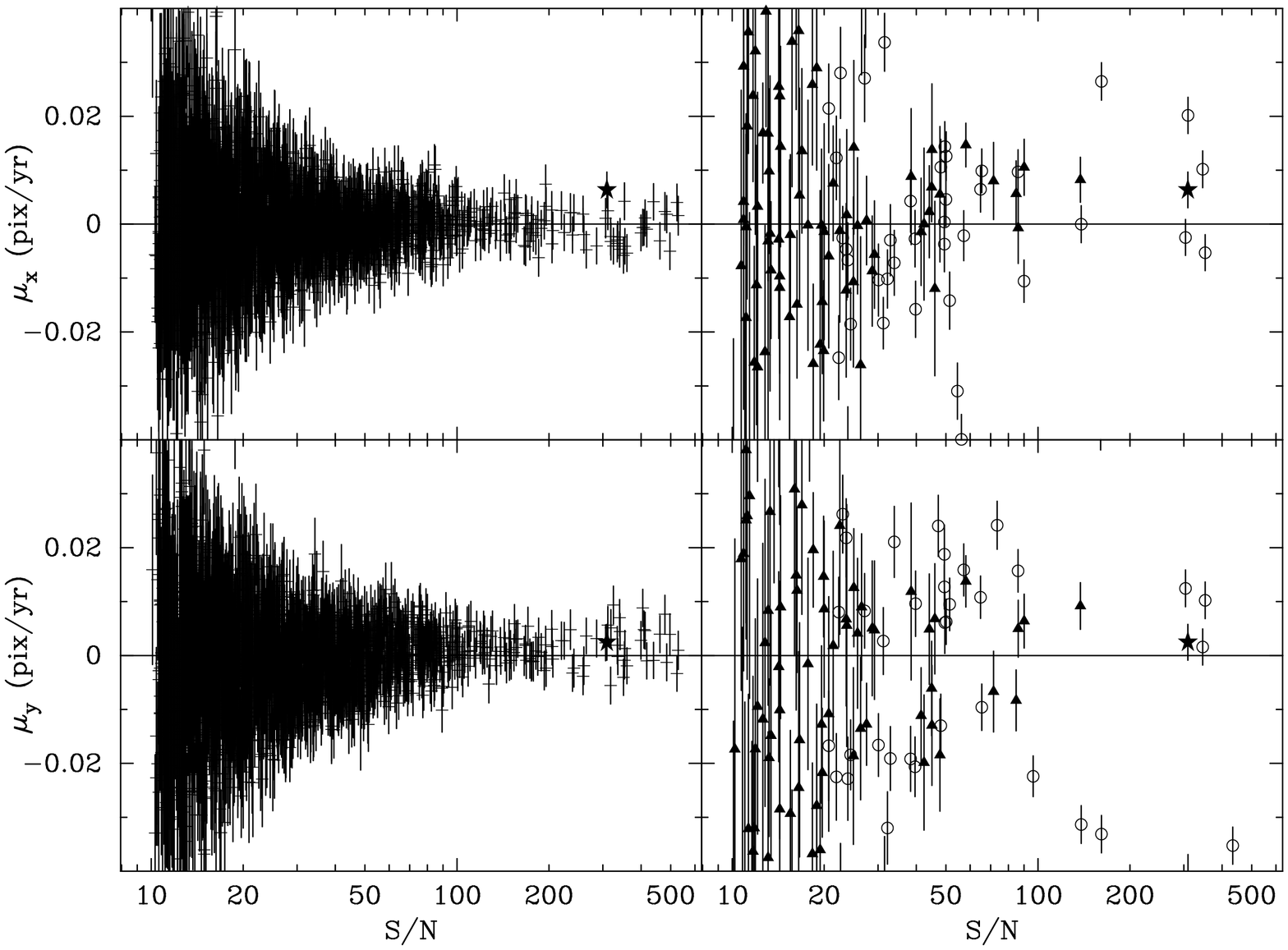}
\caption{The same as Figure~\ref{fig:mu_xy_SN_WFC1} for the WFC2 field.}
\label{fig:mu_xy_SN_WFC2}
\end{figure}

Figure~\ref{fig:mu_xy_SN_WFC1}\ shows for the WFC1 field $\mu_x$ (top
panels) and $\mu_y$ (bottom panels) as a function of S/N for (left
panels) the likely stars of Draco and the QSO and (right panels) the
background galaxies, field stars, and -- again -- the QSO. A plus
symbol represents a likely star of Draco, a filled star the QSO, a
filled triangle a background galaxy, and an open circle a field
star. Figure~\ref{fig:mu_xy_SN_WFC2} is the corresponding plot for the
WFC2 field. In both figures, the plots for the likely stars of Draco
show that their mean motion in the standard coordinate system,
$(<\mu_x>, <\mu_y>)$, is consistent with zero, albeit with a scatter
that increases with decreasing S/N. The plots for the other objects show
that there are bright field stars with values for $(\mu_x, \mu_y)$
that are, by construction, significantly different from zero. The points
for the background galaxies have larger error bars and show a greater
scatter than the stars of Draco with the same S/N. They should not show
any systematic trends with S/N but the large scatter and the scarcity of
galaxies makes this assessment difficult.

\begin{figure}[p]
\centering
\includegraphics[angle=-90,scale=0.68]{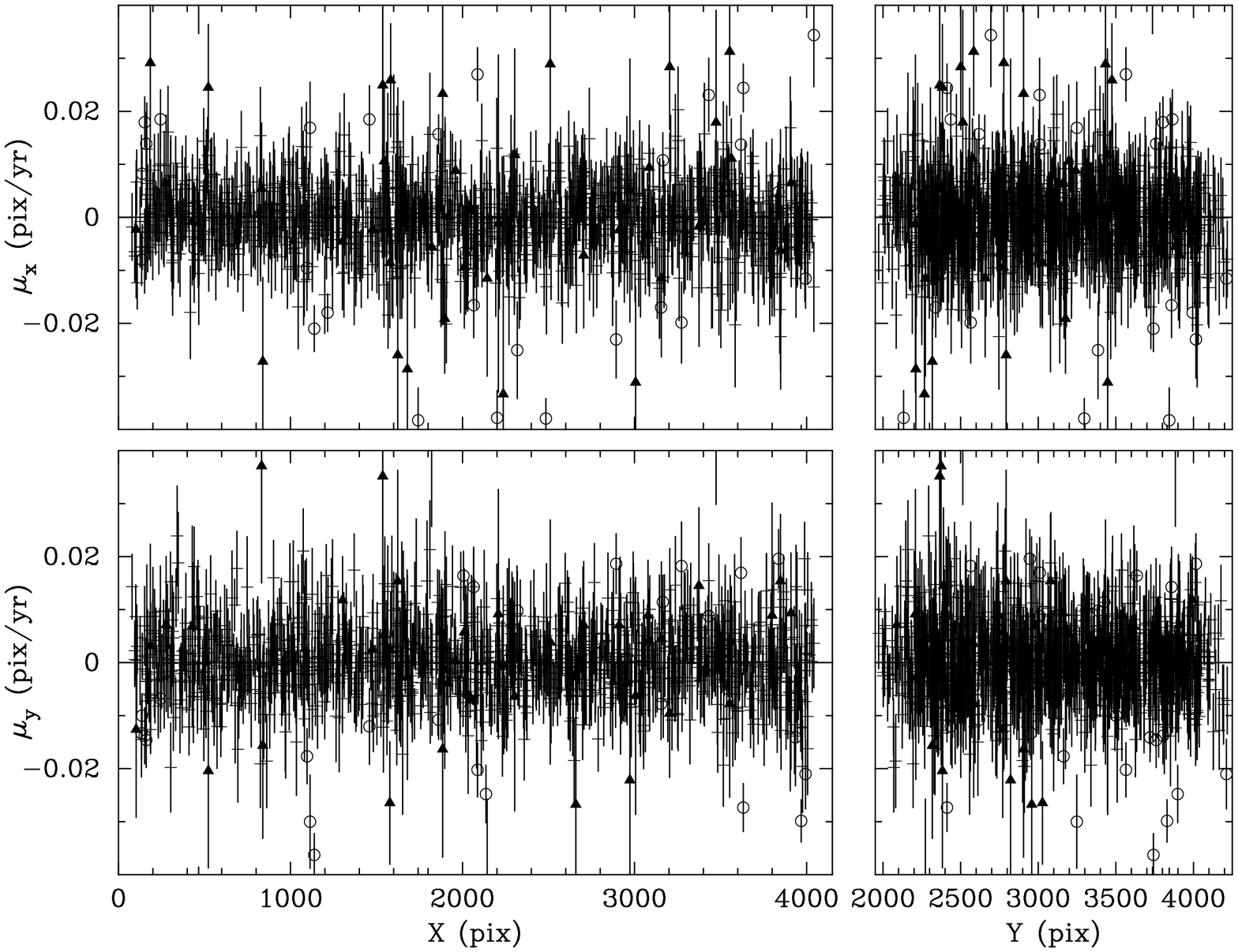}
\caption{Motions in the standard coordinate system, $\mu_{x}$ and
$\mu_{y}$, in pixel~yr$^{-1}$ as a function of $X$ (left panels) and $Y$
(right panels) for the WFC1 field. The plus symbols represent the likely
stars of Draco, filled triangles galaxies, open circles field
stars, and the filled star the QSO.  There are no obvious trends with
either $X$ or $Y$ for any class of object.}
\label{fig:mu_xy_WFC1}
\end{figure}

\begin{figure}[p]
\centering
\includegraphics[angle=-90,scale=0.68]{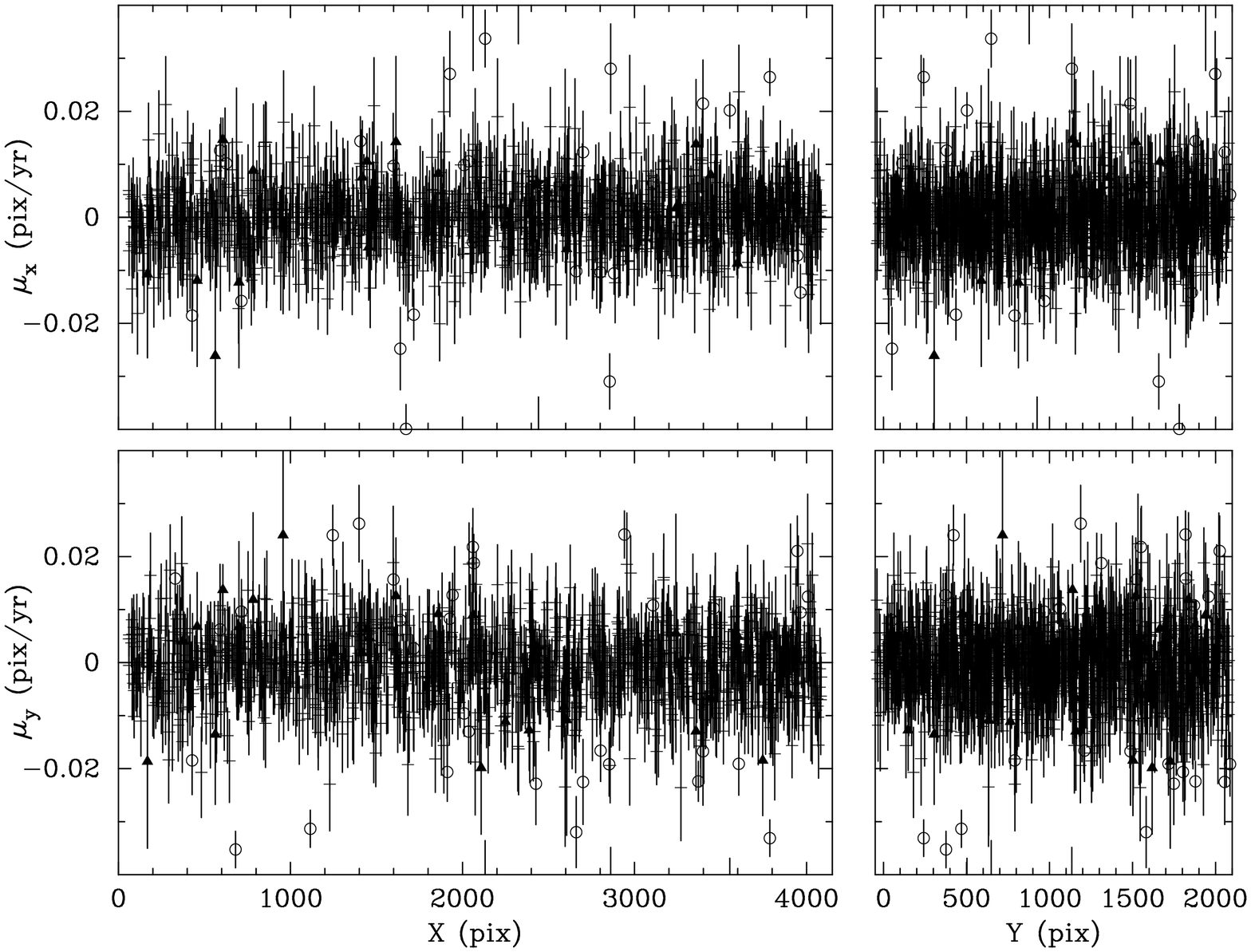}
\caption{The same as Figure~\ref{fig:mu_xy_WFC1} for the WFC2 field.}
\label{fig:mu_xy_WFC2}
\end{figure}

The four panels in Figure~\ref{fig:mu_xy_WFC1} plot $\mu_{x}$ and
$\mu_{y}$ as a function of the location $X$ and $Y$ in the detector for
the WFC1 field. Figure~\ref{fig:mu_xy_WFC2} is the same plot for the
WFC2 field. The likely stars of Draco (plus symbols) are stationary in
this system and, thus, the points should scatter around
$0$~pixel~yr$^{-1}$ and show no trends with either $X$ or $Y$, which is
the case. Although the galaxies (filled triangles) are not at rest in
the standard coordinate system, their motions should scatter around some
mean value and show no trends with either $X$ or $Y$. A visual
inspection of the plots does not show any alarming trends with position;
however, the number of galaxies is small.  Similarly, the field stars
(open circles) do not have to scatter around zero and would be unlikely
to show trends across the field.

\begin{figure}[p]
\centering
\includegraphics[angle=-90,scale=0.83]{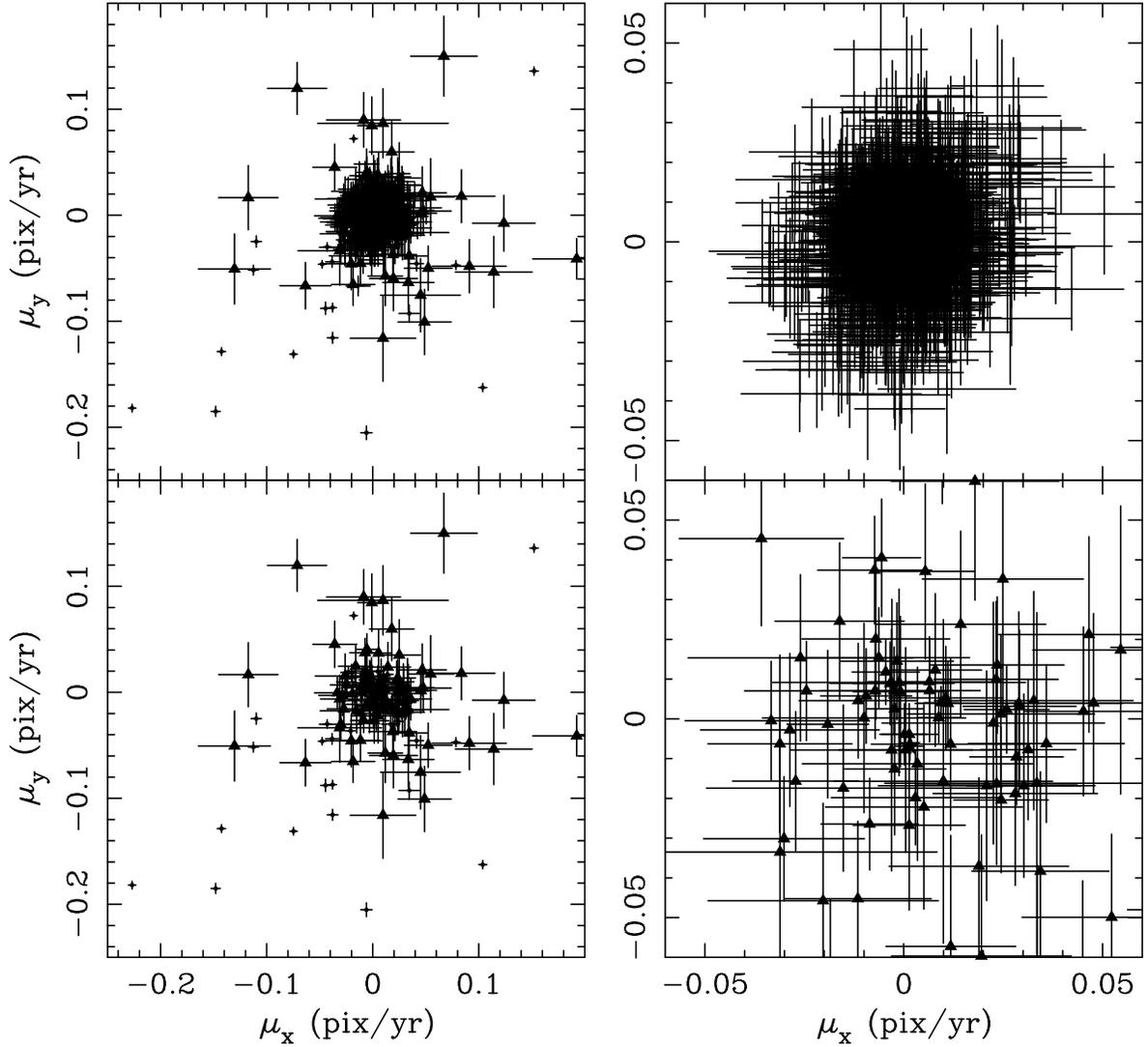}
\caption{The distribution of the measured motions in the standard
coordinate system, $(\mu_{x},\mu_{y})$, in pix~yr$^{-1}$ for the objects
in the WFC1 field.  A plus symbol represents a likely star of Draco, the
filled star the QSO, a filled triangle a galaxy, and an open circle a
field star. Top-left panel: the distribution for all of the objects.
Top-right panel: that for the likely stars of Draco. Bottom-left panel:
That for the QSO, galaxies, and background stars with the scale the same
as that in top-left panel. Bottom-right panel: that for the QSO and
galaxies with the scale the same as that in the top-right panel.}
\label{fig:mux_muy_WFC1}
\end{figure}

\begin{figure}[b!]
\centering
\includegraphics[angle=-90,scale=0.83]{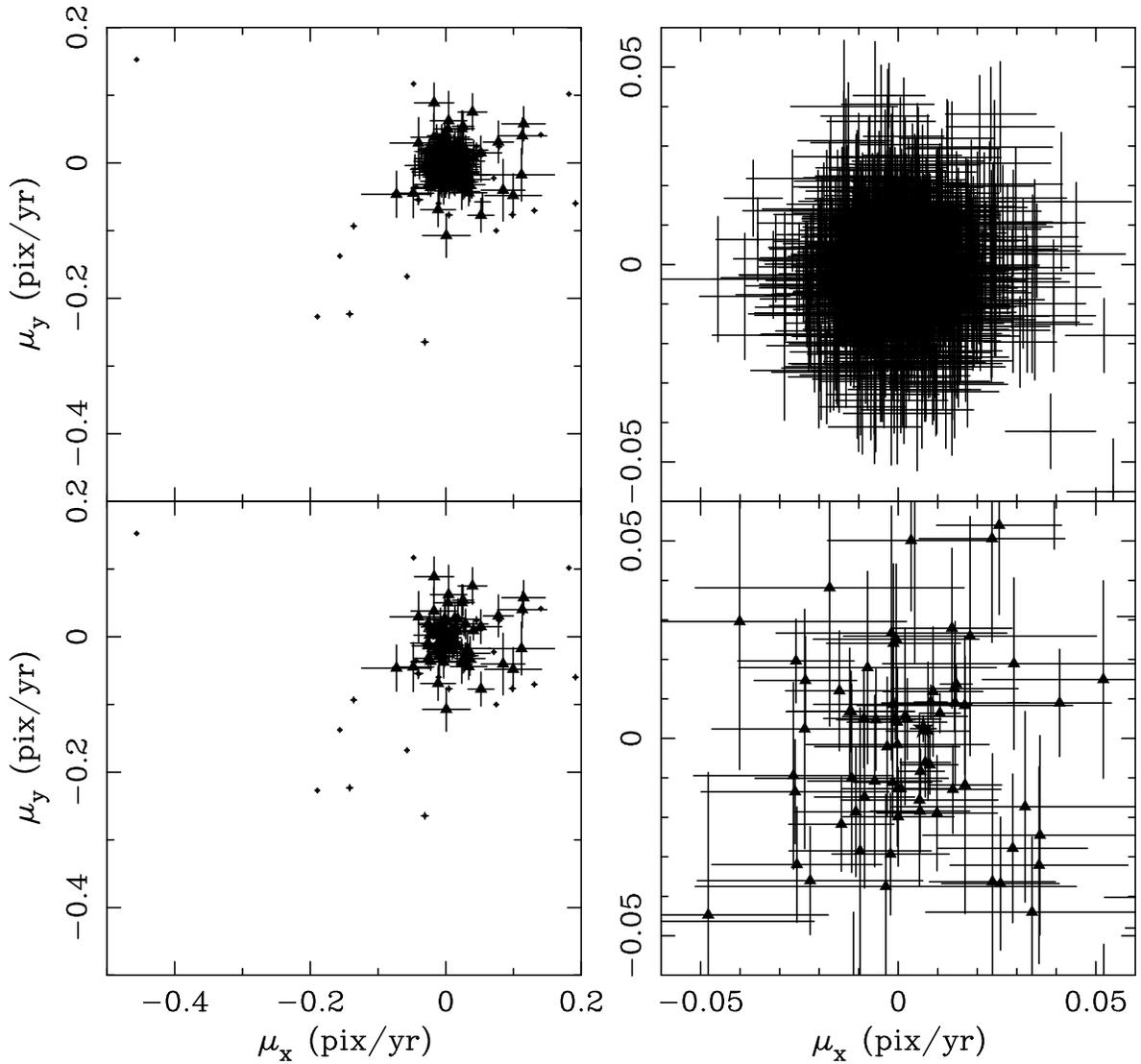}
\caption{The same as Figure~\ref{fig:mux_muy_WFC1} for the WFC2
field.}
\label{fig:mux_muy_WFC2}
\end{figure}

\newpage
        Figures~\ref{fig:mux_muy_WFC1} and \ref{fig:mux_muy_WFC2} show
the distribution of measured values of $(\mu_{x},\mu_{y})$ for the WFC1
and WFC2 fields, respectively. The symbols have the same meaning as in
the previous four figures, though the circles representing field stars
have been reduced in size to deemphasize these objects.  In each figure,
the top-left panel shows the distribtion for all of the objects (note
the different scales for the two fields). The likely stars of Draco
tightly and symmetrically clump around
$(\mu_{x},\mu_{y})=(0,0)$~pix~yr$^{-1}$. The galaxies cluster around a
point that cannot be distinguished from $(0, 0)$~pix~yr$^{-1}$ at the
scale of these plots. The field stars need not scatter around the origin
and visual inspection confirms this.  The top-right panels in
Figures~\ref{fig:mux_muy_WFC1} and \ref{fig:mux_muy_WFC2} show the
distribution for the likely stars of Draco. If only random errors are
present, the distribution should be isotropic with the density of points
increasing towards the origin; both panels show this. The bottom-left
panels depict the distribution for the galaxies and field stars. They
show more clearly the clumping of the galaxies around a point close to
the origin. Finally, the bottom-right panels are zoomed-in views of the
distribution for the galaxies only (including the QSO).

Each of the two fields yields two measurements of the proper motion of
Draco, one based on the motion of the QSO in the standard coordinate
systems and the other based on the weighted average motion of the
galaxies in those systems.  The weighted average is
$(\langle\mu_x\rangle, \langle\mu_y\rangle) =
((-5.6 \pm 2.2) \times 10^{-3}, (-0.4 \pm 2.4) \times 10^{-3})$~pixel~yr$^{-1}$
for the 97 selected galaxies in the WFC1 field and
$((-6.7 \pm 1.9) \times 10^{-3}, (-0.4 \pm 2.0) \times 10^{-3})$~pixel~yr$^{-1}$
for the 82 galaxies in the WFC2 field.  The motions for the QSOs in the
standard coordinate systems are
$((-1.6 \pm 3.6) \times 10^{-3}, (6.9 \pm 3.6) \times 10^{-3})$~pixel~yr$^{-1}$
and
$((-6.3 \pm 3.3) \times 10^{-3}, (-2.4 \pm 3.3) \times 10^{-3})$~pixel~yr$^{-1}$
for the WFC1 and WFC2 fields, respectively.  The average motion of the
galaxies and the motion of the QSO within a field are directly comparable.
Because the WFC1 and WFC2 detectors have nearly the same orientation and
have the same readout direction, the two standard coordinate systems are
nearly the same and, thus, motions in the two fields can also be compared.

The weighted average motion of the galaxies in a field does not change
significantly when the sample is reduced based on S/N.  The scatter of
the individual measured motions around their mean is greater than
expected from the uncertainties derived from those of the measured
positions at the two epochs, implying that the uncertainties in the
motions are too small.  The origin of this underestimate is difficult to
determine, therefore we have taken the empirical approach of increasing
the uncertainties in the motions by a multiplicative factor until the
$\chi^2$ of the scatter around the mean is one.  A multiplicative
increase is preferred over increasing the uncertainty by adding a
constant in quadrature because a single multiplicative factor is
sufficient for all subsamples of galaxies selected by S/N, while a different
value of the additive constant is needed for different subsamples.
The multiplicative factor is 1.55 for the WFC1 field and 1.37 for the
WFC2 field.  The uncertainties in the weighted means given above include
this increase.

\subsection{Comparison of Measurement Uncertainties}
\label{sec:puncm}

The uncertainty in each component of the motion of a QSO is about
$3.5\times 10^{-3}$~pixel~yr$^{-1}$, implying that each coordinate of
the position in the standard coordinate system has an accuracy of about
0.005~pixel at each epoch. This accuracy is comparable to the values of
0.003 -- 0.007~pixel obtained using the PC2 and STIS cameras for QSOs in
Sculptor \citep{p06} and Fornax \citep{p07}.  The fields in these
galaxies have between 200 and 500 stars, enough that uncertainties in
defining the standard coordinate system are negligible. The QSOs in the
images for Draco have a higher S/N than those for Sculptor and Fornax by
about a factor of two.  That the positional uncertainty of a QSO is
similar in all of these data argues that systematic errors are dominant
over photon statistics and, indeed, all of these studies found it
necessary to increase their uncertainties to account for such effects.

\citet{sohn13}\ measured the proper motion of Leo~I from ACS/WFC images
of a single pointing taken at two epochs separated by about 5~years
using galaxies as the astrometric zero point. Their method of deriving
the proper motion is similar, but not identical, to that used here.  The
data for Leo~I consist of fewer but deeper images per epoch than that
for Draco and, on average, have a total exposure time 1.7 times longer. 
\citet{sohn13}\ achieve an uncertainty in the proper motion that implies
the average position of the galaxies is measured with an accuracy of
0.002~pixel at both epochs, whereas the corresponding uncertainty in
the current study is 0.003~pixel.  These uncertainties imply that the
two studies are achieving comparable accuracies. 

\section{Results}
\label{sec:res}

\begin{figure}[t!]
\centering
\includegraphics[angle=-90,scale=0.8]{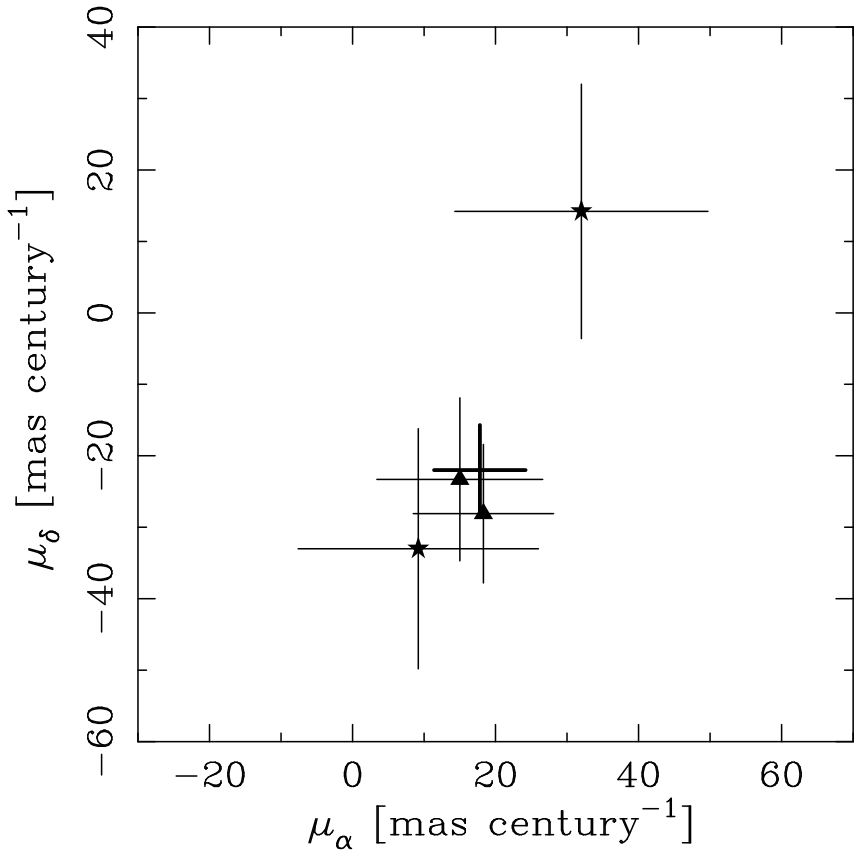}
\caption{Comparison of four estimates of the proper motion of Draco. 
The filled triangles are the values derived using the mean motions of
galaxies as the astrometric reference point, whereas the filled
stars use the QSOs.  The bold error bars show the weighted mean.
}  
\label{fig:mu_comp}
\end{figure}

Table~3 gives the four estimates of the proper motion of Draco in the
equatorial coordinate system derived from each of the mean motions of
the galaxies and of the QSOs from the previous section. These estimates
and their uncertainties are shown in Figure~\ref{fig:mu_comp}.  The
weighted average of these values is shown as bold error bars and the
value is given in the bottom line of Table~3 and Equation~\ref{eq:fmu}\
below. This measured proper motion of Draco is the main result of the
article:
\begin{equation}
(\mu_{\alpha}, \mu_{\delta}) = (17.7 \pm 6.3, -22.1 \pm 6.3)\ 
\textrm{mas}\ \textrm{century}^{-1}.
\label{eq:fmu}
\end{equation}
  
The $\chi^2$ of the scatter in the values of $\mu_{\alpha}$ is 0.99 for
three degrees of freedom.  The probability of seeing a value of $\chi^2$
this large or larger is 0.80.  Similarly, the $\chi^2$ for the values of
$\mu_{\delta}$ is 5.0, with a probability of 0.17.  The agreement of the
individual measurements with each other is acceptable and, thus, we
conclude that the uncertainties in the measured proper motion are
realistic.

The measured proper motion of Draco in the galactic coordinate system is
\begin{equation}
(\mu_{\ell}, \mu_{b}) = (-23.1 \pm 6.3, -16.3 \pm 6.3)\ 
\textrm{mas}\ \textrm{century}^{-1}.
\end{equation}
Converting the proper motion into the space velocity with
respect to the Galactic center requires adopting values for the
location and velocity of the Local Standard of Rest (LSR) and for the
velocity of the Sun with respect to the LSR.  To calculate these
``Galactic rest frame'' (Grf) quantities, we adopt a Galactocentric
radius and velocity for the LSR of 8.0~kpc
\citep[\textit{e.g.},][]{e05,gh08,gr08} and 237~km~s$^{-1}$ (consistent
with the proper motion of the Galactic center from \citet{rb04}),
respectively, and take the motion of the Sun with respect to the LSR
to be $(u_\odot,v_\odot,w_\odot) = (-10.00\pm 0.36, 5.25\pm 0.62,
7.17\pm 0.38)$~km~s$^{-1}$ \citep{db98}, where the components are
positive if $u_\odot$ points radially away from the Galactic center,
$v_\odot$ is in the direction of the rotation of the Galactic disk,
and $w_\odot$ points in the direction of the North Galactic Pole. With
these values and the heliocentric distance and radial velocity
from Table~1,
$(\mu_{\alpha}^{\mbox{\tiny{Grf}}},\mu_{\delta}^{\mbox{\tiny{Grf}}}) =
(51.4\pm 6.3, -18.7 \pm 6.3)$~mas~century$^{-1}$,
$(\mu_\ell^{\mbox{\tiny{Grf}}}, \mu_b^{\mbox{\tiny{Grf}}}) =
(-21.8\pm 6.3,-50.1\pm 6.3)$~mas~century$^{-1}$,
$(\Pi, \Theta, Z) = (27\pm 14, 89\pm 25, -212\pm 20)$~km~s$^{-1}$, and
$V_r = -98.5\pm 2.6$~km~s$^{-1}$ and $V_t = 210\pm 25$~km~s$^{-1}$.
The uncertainties in the above quantities come from Monte Carlo
experiments based on the uncertainties in the measured proper motion and
radial velocity, but not including that in the distance.  The
uncertainty in the distance from Table~1 would imply a total uncertainty
in $V_t$ of 29~km~s$^{-1}$ (determined by adding the 15~km~s$^{-1}$
contribution in quadrature).  The contribution from the uncertainty in
the distance has not been included because most of it is systematic
uncertainty in the mean absolute magnitude of RR~Lyrae variables as a
function of metallicity \citep{k08}.

The above velocities imply that Draco is moving towards the Galactic
center at the present time on an orbit with an inclination to the
Galactic plane of $70^\circ$, with a 95\% confidence interval of
$(59^\circ, 80^\circ)$. The position angle of the proper motion in the
Galactic rest frame is $110^\circ \pm 7^\circ$, which is $22^\circ$ from
the position angle of the major axis given in Table~1.  The difference
between these angles suggests that the observed elongation of Draco is
not in the plane of its orbit.  \citet{pkj13} find that Draco is part of
a planar distribution of satellite galaxies and estimate that the normal
of the plane is in the direction $(\ell, b) = (169.5^\circ, -2.8^\circ)$
with an uncertainty of $0.43^\circ$.  The velocity of Draco in our study
implies that the direction of the orbital angular momentum vector is
$(\ell,b) = (168.2^\circ\pm 4.6^\circ, -20.4^\circ\pm 5.5^\circ)$, which
is consistent with an orbit within the plane \citep{pk13}. 

\section{Summary and Discussion}
\label{sec:summary}

We have measured the proper motion of the Draco dwarf galaxy using a
single pointing imaged with ACS/WFC at two epochs separated by approximately
two years. The main conclusions of this work are as follows.

\begin{enumerate}
\item Exposure times of $19 \times 430$~s per epoch provide enough
compact galaxies in each of the WFC1 and WFC2 fields to give an
astrometric zero point whose accuracy is comparable to that given by a
single QSO.

\item The uncertainty in the proper motion achieved using the methods
in this article is comparable to that achieved by \citet{sohn13} for
Leo~I using similar data, when account is taken of the different time
baselines.

\item QSOs and background galaxies in each of the two fields all yield
consistent measurements of the proper motion.  The weighted mean of
the four values is
$(\mu_{\alpha}, \mu_{\delta}) = (17.8 \pm 6.4, -22.0 \pm 6.3)\ 
\textrm{mas}\ \textrm{century}^{-1}$.

\item The plane of the orbit of Draco is consistent with the vast polar
structure (VPOS) of Galactic satellites described by \citet{pkj13}.

\end{enumerate}

\citet{si94} used plates from the Palomar and Tautenburg Schmidt
telescopes with a time baseline of about 35~years to measure a proper
motion for Draco: $(\mu_{\alpha}, \mu_{\delta}) = (60 \pm 40, 110 \pm
50)\ \textrm{mas}\ \textrm{century}^{-1}$.  The value for $\mu_{\alpha}$
agrees with our value within the uncertainties, but the value for
$\mu_{\delta}$ differs by more than twice the uncertainty.

An important use of the proper motions of the satellite dwarf galaxies
is to provide a test of models for galaxy formation that is independent
of the observed number and luminosities of these satellites. \citet{lrl}
compared the known satellite orbits with those predicted by $\Lambda$CDM
models and found that, while the observed orbits are more circular than
predicted, the conflict is not severe.  The study finds that a decisive
test requires $V_{t}$'s accurate to about 10~km~s$^{-1}$.  Our measured
$V_t$ for Draco has an uncertainty of 25~km~s$^{-1}$.  Obtaining an
additional epoch for any of the pointings in Table~2 with $HST$ today
would increase the time baseline by at least a factor of five and
probably reduce the uncertainty in the proper motion by the same factor.
 With this additional epoch, the only obstacle to reducing the
uncertainty in $V_t$ below 10~km~s$^{-1}$ would be the uncertainty in
the distance.  A similar outcome would result from an additional epoch
for the other galaxies in \citet{lrl}, which would increase the time
baseline and reduce of the uncertainty in $V_t$ by a factor of about
seven.

\acknowledgments

CP and SP acknowledge the financial support of the Space Telescope
Science Institute through the grants HST-GO-10229 and HST-GO-11697.
EWO acknowledges support from the Space Telescope Science Institute
through the grant HST-GO-10229 and from
the National Science Foundation through the grants AST-0807498 and
AST-1313006.

\clearpage

\begin{deluxetable}{lcc}
\tablecolumns{3}
\tablewidth{0pt} 
\tablecaption{Draco at a Glance}
\tablehead{
\colhead{Quantity} &
\colhead{Value}    &
\colhead{Reference} \\
\colhead{(1)}&
\colhead{(2)}&
\colhead{(3)}}
\startdata
Right Ascension, $\alpha$ (J2000.0) &17:20:18.1&\citet{p02a} \\
Declination, $\delta$ (J2000.0) &57:55:13& $^{\prime\prime}$ \\
Galactic longitude, $\ell$& $86.3730^{\circ}$& \\
Galactic latitude, $b$    & $34.7088^{\circ}$& \\
Heliocentric distance & $82.4 \pm 5.8$~kpc& \citet{k08}\\
Position angle & $88^{\circ}\pm 3^{\circ}$ & \citet{oden01} \\
Ellipticity, $e$ &$0.29\pm0.02$& $^{\prime\prime}$ \\
Heliocentric radial velocity&$-293.3 \pm 1.0$~km~s$^{-1}$ &
\citet{aop95} \\
\enddata
\end{deluxetable}

\begin{deluxetable}{lclclll}
\tablecolumns{7}
\tablewidth{0pt} 
\tablecaption{Information about Pointings and Images}
\tablehead{    &
\colhead{R.A.} &
\colhead{Decl.}&
\colhead{Date} &
               &
               &
\colhead{T$_{exp}$}\\
\colhead{Pointing}        &
\colhead{(J2000.0)}    &
\colhead{(J2000.0)}    &
\colhead{$yyyy-mm-dd$} &
\colhead{Detector}     &
\colhead{Filter}       &
\colhead{(s)}          \\
\colhead{(1)}&
\colhead{(2)}&
\colhead{(3)}&
\colhead{(4)}&
\colhead{(5)}&
\colhead{(6)}&
\colhead{(7)}}
\startdata

Dra~1&17:19:34.4&57:58:49.8&$2004-10-30$&ACS/WFC&F555W&$19\times 430$\\
&&&$2007-11-03$&WFPC2/PC2&F555W&$12\times600$\\
\noalign{\vspace{3pt}}
Dra~2&17:20:52.3&57:55:13.4&$2004-10-19$&ACS/WFC&F606W&$19\times430$\\
&&&$2006-10-15$&ACS/WFC&F606W&$19\times427$\\
\noalign{\vspace{3pt}}
Dra~3&17:21:48.3&57:58:05.4&$2004-10-31$&ACS/WFC&F606W&$19\times430$\\
&&&$2007-12-29$&WFPC2/PC2&F606W&$12\times600$\\
\noalign{\vspace{3pt}}
\enddata
\end{deluxetable}

\begin{deluxetable}{lrr}
\tablecolumns{3}
\tablewidth{0pt}
\tablecaption{Measured Proper Motion of Draco}
\tablehead{
&\colhead{$\mu_{\alpha}$}&\colhead{$\mu_{\delta}$}  \\ 
\colhead{Field}&\multicolumn{2}{c}{(mas century$^{-1}$)}\\
\colhead{(1)}&\colhead{(2)}&\colhead{(3)}}
\startdata
WFC1 (QSO)&$32.0\pm17.7$&$14.2\pm17.8$\\
\noalign{\vspace{3pt}}
WFC1 (Galaxies)&$15.0\pm11.6$&$-23.3\pm11.4$\\
\noalign{\vspace{3pt}}
WFC2(QSO)&$9.0\pm16.5$&$-32.6\pm16.4$\\
\noalign{\vspace{3pt}}
WFC2 (Galaxies)&$18.2\pm9.6$&$-28.1\pm9.6$\\
\noalign{\vspace{1pt}}
\tableline
\noalign{\vspace{1pt}}
Weighted mean&$17.7\pm6.3$&$-22.1\pm6.3$\\
\enddata
\end{deluxetable}


\clearpage

\begin{thebibliography}{}

\bibitem[Ahn et al.(2012)]{dr9} Ahn, C. P., Alexandroff, R., Allende
Prieto, C., et al. 2012, ApJS, 203, 21
\bibitem[Anderson(2006)]{a06} Anderson, J. 2006, in The 2005 HST Calibration Workshop, ed. A. M. Koekemoer, P. Goudfrooij, \& L. Dressel (Baltimore, MD: STScI), 11
\bibitem[Anderson \& King(2000)]{ak00} Anderson, J., \& King, I. R. 2000, PASP, 112, 1360  
\bibitem[Armandroff, Olszewski \& Pryor(1995)]{aop95}Armandroff, T. E., Olszewski, E. W., \& Pryor, C. 1995, AJ, 110, 2131
\bibitem[Bertin \& Arnouts(1996)]{ba96} Bertin, E. \& Arnouts, S. 1996, A\&AS, 117, 393
\bibitem[Bristow et al.(2005)]{bpp05} Bristow, P., Piatek, S., \& Pryor,
C. 2005, ST-ECF Newsletter, 38, 12\bibitem[Dolphin(2000)]{dol} Dolphin, A. E. 2000, PASP, 112, 1383
\bibitem[Dehnen \& Binney(1998)]{db98} Dehnen, W., \& Binney, J. J. 1998, MNRAS, 298, 387
\bibitem[Eisenhauer et al.(2005)]{e05} Eisenhauer, F., et al. (2005), ApJ,
628, 246
\bibitem[Flesch(2010)]{fl10} Flesch, E. 2010, PASA, 27, 283
\bibitem[Ghez et al.(2008)]{gh08} Ghez, A. M., et al. 2008, ApJ, 689, 1044
\bibitem[Groenewegen et al.(2008)]{gr08} Groenewegen, M. A. T., Udalski, A.,
\& Bono, G. 2008, A\&A 481, 441
\bibitem[Kallivayalil et al.(2013)]{kmb13} Kallivayalil, N., van der Marel, R. P., Besla, G., Anderson, J., \& Alcock, C. 2013, ApJ, 764, 161
\bibitem[Kinemuchi et al.(2008)]{k08}Kinemuchi, K., Harris, H. C., Smith, H. A.,
Silbermann, N. A., Snyder, L. A., LaCluyz\'{e}, A. P., \& Clark, C. L. 2008, AJ, 136, 1921
\bibitem[Mahmud \& Anderson(2008)]{ma08} Mahmud, N. \& Anderson, J. 2008, PASP, 120, 907
\bibitem[Milone et al.(2006)]{mil06}Milone, A. P., Villanova, S., Bedin, L. R., et al. 2006, A\&A, 456, 517
\bibitem[L\'{e}pine et al.(2011)]{l11} L\'{e}pine, S., Koch, A., Rich, R. M., \& Kuijken K. 2011, ApJ, 741, 100L
\bibitem[Lux et al.(2010)]{lrl} Lux, H., Read, J. I., \& Lake, G. 2010, MNRAS, 406, 2312
\bibitem[Odenkirchen et al.(2001)]{oden01} Odenkirchen, M., et al. 2001, AJ, 122, 2538
\bibitem[Pawloski \& Kroupa(2013)]{pk13} Pawlowski, M. S.\& Kroupa, P. 2013, MNRAS, 435, 2116
\bibitem[Pawlowski et al.(2013)]{pkj13} Pawlowski, M. S., Kroupa, P., Jerjen, H. 2013, MNRAS, 435, 1928
\bibitem[Piatek et al.(2002a)]{p02a} Piatek, S., Pryor, C., Armandroff, T. E., \& Olszewski, E. W. 2002, AJ, 
123, 2511 
\bibitem[Piatek et al.(2002b)]{p02b} Piatek, S., Pryor, C., Olszewski, E. W., Harris, H. C.,
Mateo, M., Minniti, D., \& Tinney, C. G. 2002, AJ, 124, 3198
\bibitem[Piatek et al.(2006)]{p06} Piatek, S., Pryor, C., Bristow, P.,
Olszewski, E. W., Harris, H. C., Mateo, M., Minniti, D., \& Tinney,
C. G. 2006, AJ, 131, 1445
\bibitem[Piatek et al.(2007)]{p07} Piatek, S., Pryor, C., Bristow, P.,
Olszewski, E. W., Harris, H. C., Mateo, M., Minniti, D., \& Tinney,
C. G. 2007, AJ, 133, 818
\bibitem[Pryor et al.(2010)]{ppo10} Pryor, C., Piatek, S., \& Olszewski, E. W. 2010, AJ, 139, 839
\bibitem[Reid \& Brunthaler(2004)]{rb04} Reid, M. J., \& Brunthaler, A. 2004, ApJ, 616, 872
\bibitem[Richards et al.(2009)]{r09} Richards G.T. Myers A.D., Gray A. G., et al. 2009, ApJS, 180, 67
\bibitem[Sohn et al.(2010)]{s10} Sohn, S. T., Anderson, J., \& van der Marel, R. P. 2010, in ``The 2010 STScI 
Calibration Workshop: Hubble after SM4. Preparing JWST'' Edited by Susana Deustua and Cristina Oliveira, p. 294
\bibitem[Sohn et al.(2012)]{sav12} Sohn, S. T., Anderson, J., \& van der Marel, R. P. 2012, ApJ, 753, 7S
\bibitem[Sohn et al.(2013)]{sohn13} Sohn, S. T., Besla, G., van der Marel, R. P., Boylan-Kolchin, M., Majewski, S., \& Bullock, J. S. 2013, ApJ, 768, 139
\bibitem[Scholz \& Irwin(1994)]{si94} Scholz, R.-D., \& Irwin, M. J. 1994, \textit{Astronomy from Wide-Field Imaging} (IAU Symp.\ 161), ed. H. T. MacGillivray et~al.\ (Dordrecht: Kluwer), 534
\bibitem[Stetson(1987)]{st87} Stetson, P. B. 1987, PASP, 99, 191
\bibitem[Stetson(1992)]{st92} Stetson, P. B. 1992, in ASO Conf. Ser. Vol. 25, 
Astronomical Data Analysis Software and Systems, ed. D. M. Worrall,
C. Biemesderfer, \& J. Barnes (San Francisco: ASP), 297
\bibitem[Stetson(1994)]{st94} Stetson, P. B. 1994, PASP, 106, 250

\end{thebibliography}
\end{document}